\documentclass[12pt]{iopart}
\usepackage[dvips]{graphicx}
\usepackage{amssymb}
\usepackage{amsfonts}

\begin{document}

\bibliographystyle{iopart-num}

\title[Phonon dispersions of cluster crystals]
{Phonon dispersions of cluster crystals}

\author{Tim Neuhaus$^1$ and
Christos N Likos$^{2}$\footnote{Corresponding author, E-mail address: christos.likos@univie.ac.at}}

\address{$^1$ Institute of Theoretical Physics, Heinrich Heine University 
of D{\"u}ssedorf, Universit{\"a}tsstra{\ss}e 1, D-40225 D{\"u}sseldorf, Germany}
\address{$^2$ Faculty of Physics, University of Vienna, 
Boltzmanngasse 5, A-1090 Vienna, Austria}

\begin{abstract}
We analyze the ground states and the elementary collective excitations
(phonons) of a class of systems, which form cluster crystals in the
absence of attractions. Whereas the regime of moderate-to-high-temperatures
in the phase diagram has been analyzed in detail by means of 
density functional considerations (Likos C N, Mladek B M,
Gottwald D and Kahl G 2007 {\it J.~Chem.~Phys.}\ {\bf 126} 224502),
the present approach focuses on the complementary
regime of low temperatures. We establish the existence of an infinite
cascade of isostructural transitions between crystals with different
lattice site occupancy at $T=0$ and we quantitatively demonstrate that
the thermodynamic instabilities are bracketed by mechanical instabilities
arising from long-wavelength acoustical phonons. We further show that
all optical modes are degenerate and flat, giving rise to perfect realizations
of Einstein crystals. We calculate analytically the complete phonon spectrum
for the whole class of models as well as the Helmholtz free energy of the
systems. On the basis of the latter, 
we demonstrate that the aforementioned isostructural phase transitions
must terminate at an infinity of critical points at low temperatures,
brought about by the anharmonic contributions in the
Hamiltonian and the hopping events in the crystals.
\end{abstract}


%\maketitle

\section{Introduction}
\label{intro:sec}
Particles interacting by means of bounded 
and purely repulsive interaction potentials $v(r)$
can form cluster crystals at sufficiently high 
densities \cite{likos:pre:01,mladek:prl:06,likos:jcp:07,mladek:jpc:07,primoz:epl:08,
suto:prl:05,suto:prb:06,likos:nature:06,sven:sm:09}. 
Though the
existence of clustering in the full absence of attractions might seem
counterintuitive at first, its existence rests on solid mathematical
and physical grounds and has also been amply demonstrated by means
of detailed computer 
simulations \cite{mladek:prl:06,daan:prl:07,bianca:jpcm:08}. 
A necessary and sufficient condition
for the occurrence of clustering is that the Fourier transform of 
the interaction potential, $\tilde v(k)$, has negative parts. In this
case, the properties of the system, both in the liquid and in the
crystal phases, are largely determined by the position of the
wavevector, $k_*$, at which $\tilde v(k)$ attains its most negative
value and by the negative amplitude $-|\tilde v(k_*)|$ of the Fourier
spectrum of the potential there \cite{likos:jcp:07}. The physical realizability of such
potentials as {\it effective interactions} between suitably tailored
macromolecules has been demonstrated for the case of amphiphilic
dendrimers as well as ring polymers \cite{bianca:prl:08,arturo:sm:10}.

At moderate to high temperatures, the thermodynamics of the system
is very accurately described by a mean-field density functional theory,
which predicts, among other properties, that the lattice constant of
the crystal is density-independent \cite{likos:jcp:07}. 
This is brought about by the mechanism
of occupying the crystal sites by a multiple number of particles $n_c$
which scales proportionally to the density of particles, $\rho$, in the
crystal. Within this crystal, the equilibrium dynamics of the particles
is characterized by two kinds of processes: the first one, operating
at short time scales, is determined by lattice oscillations. At the
same time, there is incessant particle hopping between sites, which
brings about an overall average occupancy $n_c$ per site. The latter
process is activated, diffusive dynamics, which brings about a self-diffusion
coefficient of the particles that scales as $\exp(-\psi\rho/T)$, 
where $T$ is the absolute temperature and $\psi$ is a coefficient that
depends on the precise form of the interaction potential \cite{angel:prl:07}.

Whereas the diffusive, long-time dynamics of the clustered crystals is
by now well-understood due to a number of different studies, the phonon
dynamics, which is the dominant one at very low temperatures, has
not been analyzed in detail. Similarly, whereas the phase behavior
of the system at moderate to high temperatures is very thoroughly
examined, little is known regarding the phase diagram of the system
at very low temperatures. The analysis of the elementary excitations
(phonons) in the cluster crystals opens a route to 
low-temperature thermodynamics as well, since it allows for the calculation
of the free energy of the system on the basis of a harmonic lattice theory.
The purpose of this paper is precisely to calculate and analyze the
phonon spectra of cluster crystals and to draw, on this basis, conclusions
on the stability, dynamics and thermodynamics of these systems at low
temperatures. We find that cluster crystals have very unusual, yet simple,
phonon spectra, in which all optical modes are degenerate and the optical
frequencies are independent of the wavenumber, rendering thereby these
solids into perfect Einstein crystals. Further, we discover a cascade
of mechanical instabilities in the system, initiated at the long-wavelength
limit of the low-lying acoustical modes, which trigger an avalanche
of isostructural phase transitions from crystals of occupancy $n_c$ to
crystals of occupancy $n_c + 1$, whereby $n_c$ is an integer number. 
Finally, we predict that 
the (infinitely many)
density gaps associated with the isostructural transitions at $T=0$ 
gradually close up as $T$ grows and they terminate at critical points
at low temperatures.

The rest of the manuscript is organized as follows: 
In sec.~\ref{zerot:sec} we calculate the zero-temperature phase
diagram, to establish the ground states on which the phonon spectra
have to be determined. The calculation of the phonon curves follows
then in sec.~\ref{phonon:sec} and the ensuing phonon-based 
phase diagram of the system in sec.~\ref{phdg:sec}. Finally, in 
sec.~\ref{concl:sec} we summarize and draw our conclusions.

\section{Zero-temperature phase diagram}
\label{zerot:sec}

To set up the stage and gather the key results for the
system, we commence with the definition of the relevant quantities
and a collection of hitherto known results. We consider a system
of $N$ particles in a volume $V$, which interact via a bounded
and positive interaction potential $v(r)$:
\begin{equation}
0 \leq v(r) < \infty.
\label{vofr:eq}
\end{equation}
The system is characterized by its density $\rho = N/V$ as well as
the absolute temperature $T$. Without loss of generality, we can
introduce an energy scale $\varepsilon$ and a length scale $\sigma$
that set the strength and range of the potential $v(r)$, respectively,
writing the latter in the form:
\begin{equation}
v(r) = \varepsilon\phi(r/\sigma),
\label{phi:eq}
\end{equation}
with some dimensionless function $\phi(z)$ of a dimensionless argument
$z$. Accordingly, one can define the scaled density $\rho^* \equiv 
\rho\sigma^3$ as well as the scaled temperature $T^* \equiv 
k_{\rm B}T/\varepsilon$, where $k_{\rm B}$ is Boltzmann's constant.
Another crucial quantity is the Fourier transform ${\tilde v}(k)$
of $v(r)$, which contains negative parts, i.e., $v(r)$ belongs to
what has been termed the $Q^{\pm}$ class of interactions \cite{likos:pre:01}.
In particular, we consider its most negative value, which is assumed
at the wavenumber $k_*$:
\begin{equation}
-|{\tilde v}(k_*)| = \min_{k}\,{\tilde v}(k).
\label{vofkmin:eq}
\end{equation}

At temperatures $T^* \gtrsim 0.1$, analytical considerations and
comparisons with extensive computer simulations have shown that the
system forms clustered crystals with the following 
properties \cite{mladek:prl:06,likos:jcp:07,mladek:jpc:07,daan:prl:07}. First,
the freezing line $(T_{\rm f},\rho_{\rm f})$ 
from a fluid (at low densities) to the cluster
crystal (at high densities) takes, approximately, the form:
\begin{equation}
k_{\rm B}T_{\rm f} \approx 1.393 |{\tilde v}(k_*)| \rho_{\rm f}.
\label{freeze:eq}
\end{equation}
Second, the clustered crystals feature an average lattice site
occupancy $n_c$ that scales as
\begin{equation}
n_c = \frac{8\sqrt{2}\pi^3}{k_*^3}\rho,
\label{nc:eq}
\end{equation}
resulting into a density-independent lattice constant $a$ of the crystal.
Finally, the long-time, diffusive dynamics of the crystals is determined
by an activated-hopping mechanism, resulting into a diffusivity $D$ that
is given by \cite{angel:prl:07}:
\begin{equation}
D \cong [(\psi \rho^*/T^*)^2/2 + \psi\rho^*/T^* + 1]\exp(-\psi\rho^*/T^*),
\label{diffusivity:eq}
\end{equation}
with a potential-dependent coefficient $\psi$. 

As is clear from eq.~(\ref{nc:eq}) above, the average occupancy $n_c$ of the
lattice is a general, real number. At any instant, of course, every
crystal site is occupied by some integer number of particles. At finite
temperatures, the activated hopping mechanism that brings about the
long-time diffusivity $D$ of eq.~(\ref{diffusivity:eq}), constantly
redistributes particles between instantaneous `donor' and `acceptor'
sites, bringing about the aforementioned average occupancy $n_c$.
At $T=0$, however, this mechanism is absent, therefore each site
has an occupancy that is frozen-in. In addition, the zero-temperature
phase has to fulfill the strict requirement of {\it mechanical
equilibrium}\footnote{Mechanical equilibrium is a necessary condition
for stability of a structure, not a sufficient one; the increasingly
stronger requirements of mechanical and thermodynamic stability must
also be fulfilled for a phase to be materialized.
The question of
the mechanical stability of this equilibrium will be discussed in 
sec.~\ref{phonon:sec}.} 
on each and every lattice site, i.e., the {\it force}
exercised on every particle on the lattice site, due to the rest
of the crystal must vanish. This immediately excludes the possibility
of a periodic crystal with arbitrarily varying occupancies between
the lattice sites: inversion symmetry of the decorated crystal structure
must be guaranteed, so that mechanical equilibrium can result. 

\begin{figure}
\begin{center}
\includegraphics[width=12cm,clip=true]{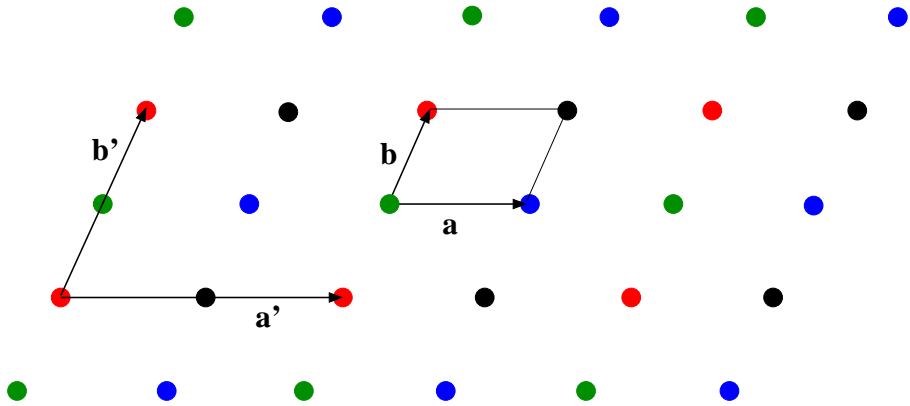}
\end{center}
\caption{A schematic demonstration of the procedure to generate the
most general decorated lattice (here in two dimensions) that fulfills
inversion symmetry (see the text). The underlying Bravais lattice
is spanned by the elementary vectors ${\bf a}$ and ${\bf b}$. The vertices
of the elementary unit cell, denoted by the parallelogram, are decorated
by clusters of different occupancy, marked by the four different colors.
Afterwards, the decorated cell is flipped along the sides generated
by the vectors ${\bf a}$ and ${\bf b}$. The procedure is repeated
for every new decorated unit cell until the whole lattice is filled.}
\label{decorate:fig}
\end{figure}

The most
general way of achieving this is depicted in Fig.~\ref{decorate:fig}.
One takes the elementary unit cell of the lattice and decorates
each vertex with (in general different) integer occupancies of 
the lattice sites. Then, the cell is `flipped' along the axes
defined by the elementary vectors and the process is repeated
until the whole lattice is filled. 
For the two-dimensional case shown 
for purposes of clarity in Fig.~\ref{decorate:fig}, the resulting
structure can also be regarded as a crystal with
a unit cell spanned by the vectors ${\bf a'} = 2{\bf a}$ and
${\bf b'} = 2{\bf b}$, accompanied by a four-membered basis at the
vectors ${\bf B}_1 = 0$, ${\bf B}_2 = {\bf a}$, 
${\bf B}_3 = {\bf b}$, and ${\bf B}_4 = {\bf a}+{\bf b}$.
The generalization to three dimensions is straightforward, where one
obtains up to eight sublattice occupancies 
$n_i$, $i=1,2,\ldots,8$ and the corresponding overall occupancy
of the lattice is a {\it rational} number:
\begin{equation}  
n_c = \frac{1}{8}\sum_{i=1}^8 n_i.
\label{ncrat:eq}
\end{equation}
Evidently, even this combinatorial freedom does not restore the
general possibility of real (i.e., in general irrational) $n_c$ numbers
present for $T \ne 0$. For the concrete example of the interaction
potential examined in the rest of the paper, we checked a few
possible combinations of $n_i$'s at selected densities and we found
that the most stable lattice decoration at $T=0$ is the one in which 
{\it all} $n_i$'s coincide. On these grounds, and on the basis
of parsimony and clarity, we will consider in the rest of the paper
precisely only the case for which every lattice site is occupied
by the same, integer number of particles. Thus, we will be discussing
crystals of single, double, triple, etc.~occupancy.

Though a host of results to be derived below are quite general for 
all potentials in the $Q^{\pm}$ class, we are working with the
generalized exponential model with exponent 4, GEM4, described
by the interaction potential:
\begin{equation}
v(r) = \varepsilon\,\exp[-(r/\sigma)^4].
\label{gem4:eq}
\end{equation}
At $T=0$, the stable crystal lattice for this potential is fcc,
as confirmed earlier by applying a very efficient search among
all Bravais lattices via a genetic algorithm 
approach \cite{mladek:prl:06,dieter:ga:05}. The ground-state
phase diagram is determined by the configurations that minimize the
internal energy $U$ (the lattice sum). In particular, 
one has to minimize the internal
energy density $U/V$ and the latter must be a convex function
of the particle density $\rho$ of the system.

\begin{figure}
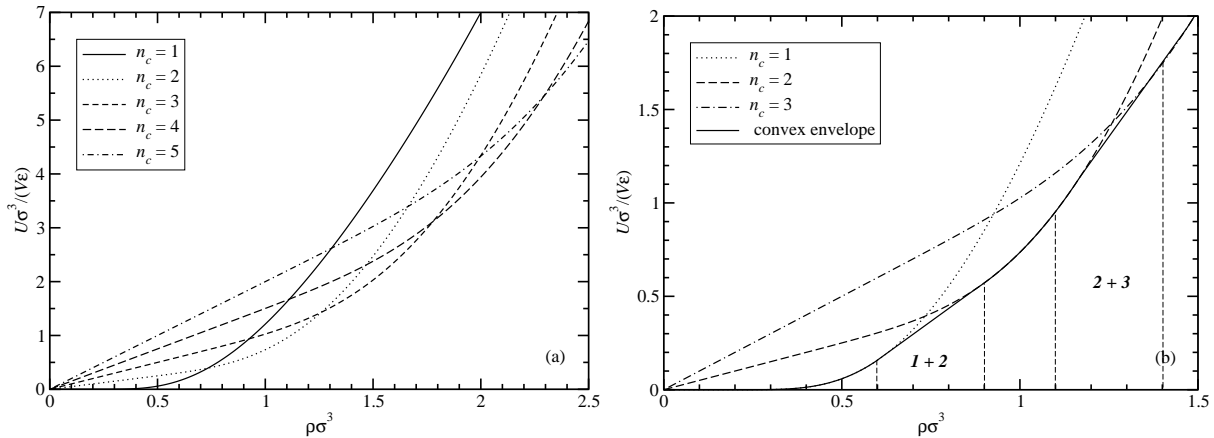

   \begin{center}
   \begin{minipage}[b]{16.0cm}
         \includegraphics[width=7.9cm,clip]
         {./fig2a.eps}
         \includegraphics[width=8.0cm,clip]
         {./fig2b.eps}
   \end{minipage}
   \end{center}
\caption{(a) The internal energy density $U/V$ of GEM4 fcc crystals with
successively high occupancies per lattice site, $n_c$, as a function of
the density $\rho$. The occupancies are indicated in the legend. (b) A zoom
on the lower density part of (a), showing also the convex reconstruction
of the curves and the resulting two-phase coexistence regions between
crystals with successive occupancy numbers. The coexistence regions correspond
to the straight-line segments of the $U/V$ vs.~$\rho$-curves and are also
marked by the vertical broken lines. The inscriptions in these regions
denote the occupancies of the coexisting crystals.}
\label{groundstate:fig}
\end{figure}

Results are shown in Fig.~\ref{groundstate:fig}. In Fig.~\ref{groundstate:fig}(a)
it can be seen that with increasing density, crystals with occupancy $n_c$
become thermodynamically less favorable and they give their place to
crystals with occupancy $n_c+1$. At the same time, regions of non-convexity
appear around the points on which the $U/V$ vs.~$\rho$-curves of the
various occupancies cross. These have to be made convex by means of the
common tangent construction, as shown in Fig.~\ref{groundstate:fig}(b),
giving rise to regions of coexistence between the crystal of occupancy
$n_c$ and the crystal of occupancy $n_c+1$. 
In what follows, we denote
with $\rho_{n_c/n_c+1}^{-}$ and $\rho_{n_c/n_c+1}^{+}$ the lower and
upper density boundaries of the coexistence region between the 
fcc solids of occupancy $n_c$ and $n_c+1$. 
For consistency in the notation, let us also define $\rho_{0/1}^{+} = 0$.
The phase behavior of the system at $T=0$ can be summarized as follows.
For
$\rho_{n_c-1/n_c}^{+} \leq \rho^* \leq \rho_{n_c/n_c+1}^{-}$ a single crystal
of occupancy $n_c$ is stable, whereas for
densities $\rho^*$ in the range
$\rho_{n_c/n_c+1}^{-} < \rho^* < \rho_{n_c/n_c+1}^{+}$,
a macrophase separation between the crystals of successive occupancies
$n_c$ and $n_c+1$ takes place.

\begin{figure}
\begin{center}
\includegraphics[width=12cm,clip=true]{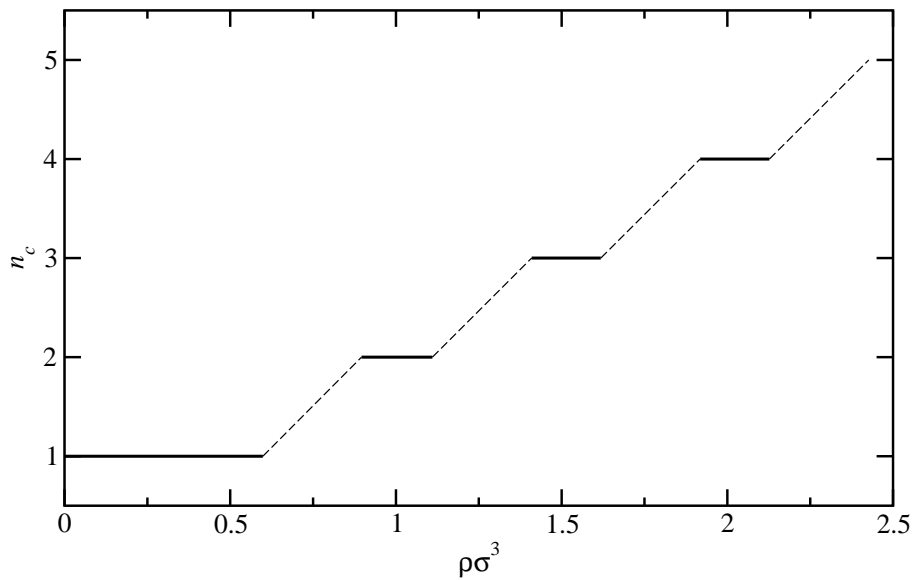}
\end{center}
\caption{The zero-temperature dependence of the lattice occupancy
$n_c$ of GEM4 fcc crystals on their density. The thick horizontal
lines lie in single-phase density regions, separated by phase-coexistence
density gaps, denoted by the thin broken lines.}
\label{occupancy_steps:fig}
\end{figure}

In Fig.~\ref{occupancy_steps:fig} we show the dependence of the 
$T=0$ site occupancy $n_c$ on density. The characteristic step-like
structure that emerges is a signature of the ground state and a
precursor of the $n_c \propto \rho$ dependence, eq.~(\ref{nc:eq}),
encountered for higher temperatures. Indeed, the `smoothing-out'
of the steplike form of the occupancy into a continuous, straight line
is brought about by the hopping mechanism mentioned above. The
ranges of stability of the $n_c$-occupied crystals and the
associated regions of coexistence for the fcc GEM4-crystals
at zero temperature are summarized in Table \ref{phdg:tab} for the
first few values of $n_c$.

\Table{\label{phdg:tab} The density ranges of stability of the 
$n_c$-occupied fcc GEM4 crystals, shown here up to the coexistence
region between quintuple and sextuple occupied solids. 
The symbols `$a + b$' on the
left column, where $b = a+1$, denote, as in the figures, the
regions of macroscopic coexistence (phase separation) between 
crystals of occupancy $n_c$ and $n_c + 1$.}
\br
Site occupancy $n_c$&Density range $\rho^*$\\
\mr
  1 & 0 - 0.5985 \\
  1 + 2 & 0.5985 - 0.8961 \\
  2 & 0.8961 - 1.1100 \\
  2 + 3 & 1.1100 - 1.4094 \\
  3 & 1.4094 - 1.6182 \\
  3 + 4 & 1.6182 - 1.9179 \\
  4 & 1.9179 - 2.1267 \\
  4 + 5 & 2.1267 - 2.4267 \\ 
  5 & 2.4267 - 2.6336 \\
  5 + 6 & 2.6336 - 2.9336 \\ 
\br
\endTable

\begin{figure}
\begin{center}
\includegraphics[width=12cm,clip=true]{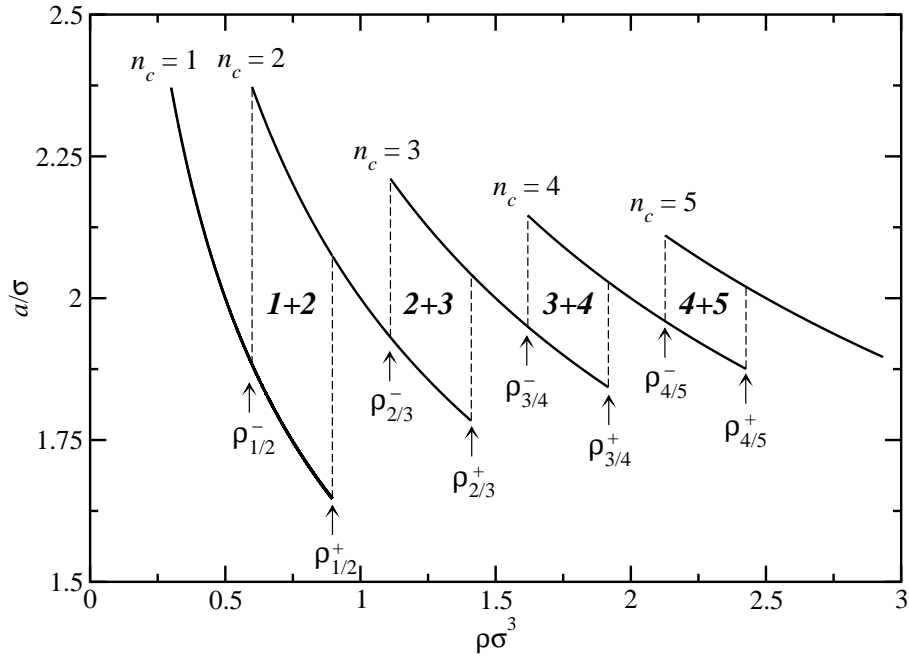}
\end{center}
\caption{The zero-temperature dependence of the lattice constant
$a$ of GEM4 fcc crystals on their density. The thick solid
lines have the dependence $a = \gamma(n_c/\rho)^{1/3}$ and they
are artificially extended beyond the range of stability of single phases,
within the coexistence regions, for purposes of
demonstration. The coexistence regions are denoted by thin
vertical lines and the inscriptions in those denote the occupancies
of the two coexisting crystals. The lower- and upper densities
of the coexistence gaps are marked with arrows. For the overall,
reconstructed dependence of $a$ on density, taking phase coexistences
properly into account, see the text.}
\label{lattice_constant:fig}
\end{figure}

In Fig.~\ref{lattice_constant:fig}, we show the resulting dependence
of the lattice constant $a$ of the fcc crystals on their density.
Within the regions $\rho_{n_c-1/n_c}^{+} \leq \rho^* \leq \rho_{n_c/n_c+1}^{-}$,
in which a pure crystal with occupancy $n_c$ is stable, the lattice constant
decreases with density as $a = \gamma(n_c/\rho)^{1/3}$, with some
geometric prefactor $\gamma$. In the regions
$\rho_{n_c/n_c+1}^{-} < \rho^* < \rho_{n_c/n_c+1}^{+}$, we have the
coexistence between two crystals, one with occupancy $n_c$ and lattice
constant $a^{-} = \gamma(n_c/\rho_{n_c/n_c+1}^{-})^{1/3}$, and one
with occupancy $n_c+1$ and lattice constant
$a^{+} = \gamma((n_c+1)/\rho_{n_c/n_c+1}^{+})^{1/3}$. Whereas for a crystal
of single occupancy the lattice constant would monotonically decrease
with density as $\rho^{-1/3}$, here we see already at $T=0$ the
propensity of the cluster crystals to suppress the variation 
of $a$ with $\rho$. This comes about through the growth of $n_c$
with $\rho$ in two complementary fashions: on the one hand, for the
single phase regions, the relative difference between 
$(n_c/\rho)^{1/3}$ and $((n_c+1)/\rho)^{1/3}$ shrinks as $n_c$ grows.
On the other hand, the intervening two-phase coexistence regions
have the effect that, although the lattice constant decreases with
$\rho$ within a single-phase region, it gets `kicked up' again
when the transition to the next single-phase region takes place.
Indeed, as can be seen in Fig.~\ref{lattice_constant:fig}, the range
of variation of $a$ with density becomes increasingly narrower as
$\rho$, and thus also $n_c$ grow. Once again, this is the $T=0$
precursor of the ensuing density-independence of $a$ on $\rho$ that
has been established for higher temperatures. The way in which the
discontinuous change of $n_c$ and/or $a$ with density 
at zero temperature evolve into the smooth curves at higher temperature
will be discussed in sec.~\ref{phdg:sec}.

\section{Phonon spectra and stability analysis}
\label{phonon:sec}

\subsection{Single occupancy crystals}
\label{acoustical:sec}

The calculation of the phonons spectra, once the ground states have
been determined, follows along standard ways \cite{am}. The particles are
subjected to small deviations around their equilibrium
lattice positions $\{{\bf R}\}$,
denoted by the displacement field ${\bf u}({\bf R})$, and the
total interaction energy is split into the lattice sum $U$ and
a harmonic potential $U^{\rm harm}$, which is a 
quadratic
form of ${\bf u}({\bf R})$:
\begin{equation}
U^{\rm harm} = \frac{1}{2}\sum_{\bf R}\sum_{\bf R'}
{\bf u}({\bf R}){\bf D}({\bf R}-{\bf R'}){\bf u}({\bf R'}),
\label{harmonic:eq}
\end{equation}
where the dynamical matrix ${\bf D}({\bf R}-{\bf R'})$ has the 
elements
\begin{equation}
D_{\alpha\beta}({\bf R}-{\bf R'}) = 
\frac{\partial^2 U^{\rm harm}}
{\partial u_{\alpha}({\bf R})\partial u_{\beta}({\bf R'})},
\label{dofr:eq}
\end{equation}
$\alpha,\beta$ denoting Cartesian coordinates. Determination of the
phonon spectrum $\omega({\bf k})$ results from diagonalization of
the matrix
\begin{equation}
{\bf D}({\bf k}) = \sum_{\bf R}{\bf D}({\bf R}) \exp[-{\rm i}{\bf k}\cdot{\bf R}].
\label{dofk:eq}
\end{equation}
In particular, the eigenvalues 
of the matrix ${\bf D}({\bf k})$ are equal to $m\omega^2({\bf k})$,
$m$ being the particle mass. 

The phonon spectra have been calculated along the  
$\Gamma \to {\rm X} \to {\rm W} \to {\rm K} \to \Gamma$-path 
in the first Brillouin
zone of the fcc-lattice, shown in Fig.~22.13(b) of Ref.~\cite{am}.
We note, in particular, that the $\Gamma \to {\rm X}$ section of this
path corresponds to wavevectors ${\bf k}$ along the $[100]$ 
direction of the cubic crystal, whereas the last part of the
path, ${\rm K} \to \Gamma$, to wavevectors along the $[110]$ direction,
which is also the nearest-neighbor vector in the fcc crystal.
The points ${\rm X}$ and ${\rm K}$ lie at the edges, whereas
the point $\Gamma$ at the center of the Brillouin zone.

\begin{figure}
\begin{center}
\includegraphics[width=12cm,clip=true]{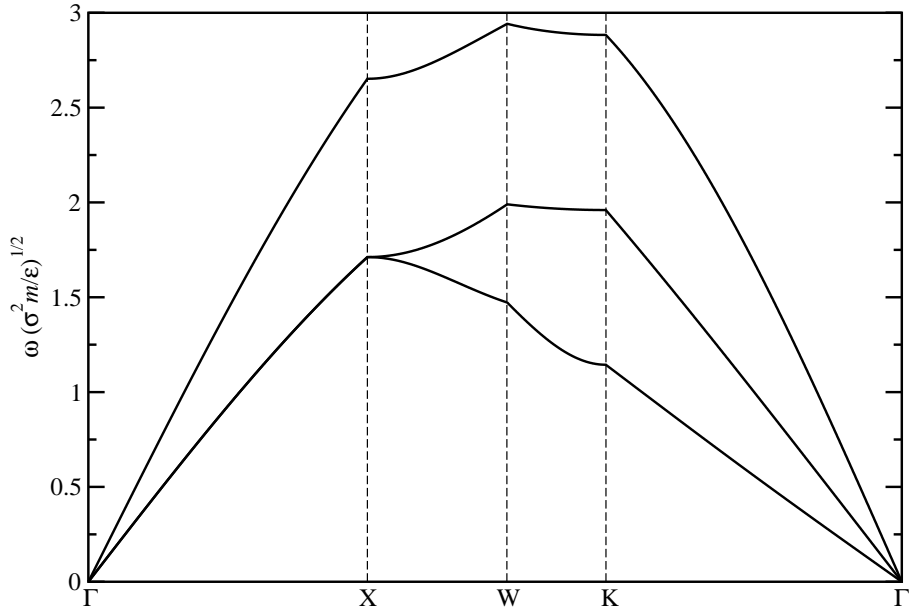}
\end{center}
\caption{The phonon spectrum of the single occupied fcc GEM4-crystal
at density $\rho^* = 0.5$.}
\label{phonons_3d_single_stable:fig}
\end{figure}

The phonon spectrum of the single occupied fcc GEM4-crystal at 
density $\rho^* = 0.5$ is shown in Fig.~\ref{phonons_3d_single_stable:fig}.
Along the high-symmetry, $\Gamma \to {\rm X}$, $[100]$ direction,
we find, as expected, two degenerate transverse acoustical branches,
as well as a separate longitudinal branch. The degeneracy of the
two former ones is lifted along the other directions in the Brillouin
zone. Upon increasing the density of the crystal, starting at 
vanishingly small values, the slope of the lowest branch in the
$\Gamma \to {\rm K}$ direction has a nonmonotonic behavior. It first
increases and then starts decreasing again, a development that is
at odds with that for usual, monatomic crystals. The decrease of
the slope results, at a density $\rho^*_{\times, 1}$ at the appearance
of a {\it negative eigenvalue}, which corresponds to an imaginary
frequency and thus signals a {\it mechanical instability} of the
crystal. We have numerically found for $\rho^*_{\times, 1}$ the value
\begin{equation}
\rho^*_{\times, 1} = 0.7205.
\label{rhox1:eq}
\end{equation}

\begin{figure}
\begin{center}
\includegraphics[width=12cm,clip=true]{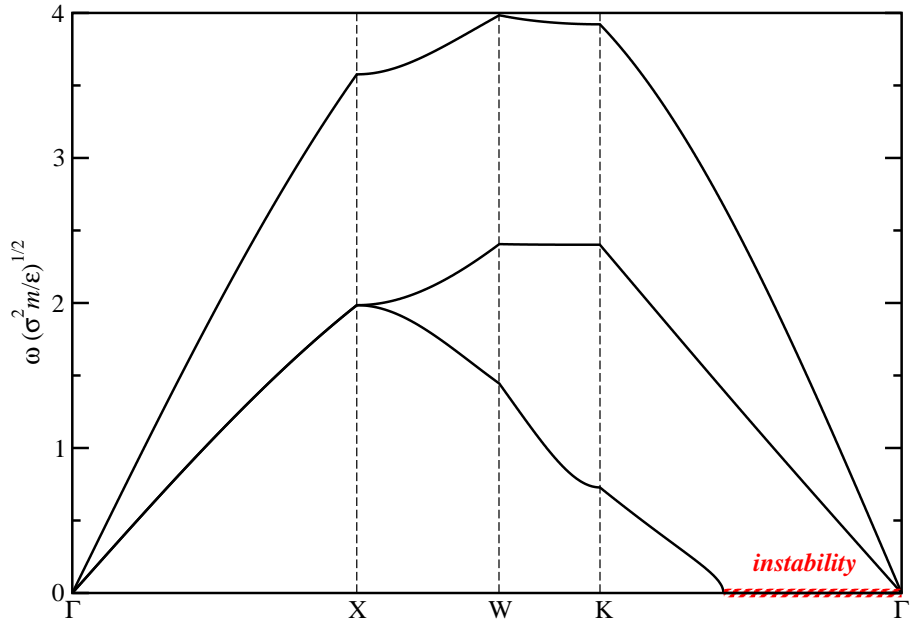}
\end{center}
\caption{The phonon spectrum of the single occupied fcc GEM4-crystal
at density $\rho^* = 0.75$. The cross-hatched region on the horizontal
axis denotes the values of wavenumbers for which the matrix
${\bf D}({\bf k})$ has negative eigenvalues.}
\label{phonons_3d_single_unstable:fig}
\end{figure}

Since the instability first occurs at the point $\Gamma$, i.e., for
$k = 0$, it signals a propensity of the single occupied crystal to
undergo a long-wavelength transformation (compression/expansion),
fully consistent with the thermodynamic instability towards the
formation of a doubly occupied crystal with a different lattice
constant. Referring to Table \ref{phdg:tab}, we see that
the point of mechanical instability lies above the point of thermodynamic
instability, $\rho^{-}_{1/2} = 0.5985$, of the single occupied crystal. 
In other words, the incipient mechanical instability of the system,
at which the slightest perturbation to the crystal will result into
its collapse, is {\it preempted} by a first-order transition to a
doubly occupied crystal. The scenario is reminiscent of the 
first-order freezing transition of the cluster fluid into the
cluster crystal preempting the instability line at $k = k_*$, encountered
for the very same systems \cite{likos:pre:01,mladek:prl:06,likos:jcp:07},
the difference being that the latter occurs
at finite values of the wavenumber and signals thus an ordering 
with finite wavelength. At any rate, the fact that the thermodynamic
phase transition precedes the mechanical instability is the only
consistent scenario, since the opposite would lead to an obvious
contradiction. Upon further increasing the density beyond the value
$\rho^*_{\times, 1}$, an increasingly long segment of modes along
the $\Gamma \to {\rm K}$ branch goes unstable, as shown in 
Fig.~\ref{phonons_3d_single_unstable:fig}. 

\subsection{Multiple occupancies and the optical branches}
\label{optical:sec}

A multiply occupied crystal is a particular case of a Bravais lattice 
with a $n_c$-component basis, for which all basis vectors
${\bf B}_i$, $i=1,2,\ldots,n_c$ are equal to zero. The displacement 
field ${\bf u}({\bf R})$ has to be augmented to a `supervector'
$\{{\bf u}^{(\mu)}({\bf R})\}$, $\mu = 1,2,\ldots,n_c$, where
the superscript characterizes the particle species (all species
and their interactions
being identical in our case). The dynamical matrix in $d$ spatial
dimensions now becomes $dn_c \times dn_c$ and we obtain,
accordingly, $d$ acoustical and $d(n_c-1)$ optical branches
in the phonon spectrum.

For purposes of simplicity and transparency, we consider first,
the case $d=1$, i.e., a one-dimensional crystal of site
occupancy $n_c$, formed
by $Q^{\pm}$-particles. Due to the combination
of the facts that all interactions between the species are identical
{\it and} that all basis vectors vanish, it is straightforward to
show that the $n_c \times n_c$
dynamical matrix ${\bf D}(k)$ takes, in this case,
a special form of a Toeplitz matrix: all its diagonal 
elements are equal to $A(k)$
and all non-diagonal ones to $B(k) \ne A(k)$, namely: 
\begin{equation}
{\bf D}(k)  = \left({\begin{array}{cccccccc}
A(k) & B(k) & B(k) & B(k) & . & .    & .    & B(k)\\
B(k) & A(k) & B(k) & .    & . & .    & .    & B(k)\\
B(k) & B(k) & A(k) & .    & . & .    & .    & B(k)\\
B(k) & .    & .    & .    & . & .    & .    & .   \\
.    & .    & .    & .    & . & .    & .    & .   \\
B(k) & B(k) & B(k) & .    & . & .    & A(k) & B(k)\\
B(k) & B(k) & B(k) & .    & . & .    & .    & A(k)\end{array} } \right),
\label{matrix1d:eq}
\end{equation}
where
\begin{eqnarray}
A(k) & = & \sum_{R}v''(R)[n_c - \cos (kR)]
\label{a1d:eq}
\\
B(k) & = & -\sum_{R}v''(R)\cos (kR), 
\label{b1d:eq}
\end{eqnarray}
and the summations in eqs.~(\ref{a1d:eq}) and 
(\ref{b1d:eq}) are carried over all Bravais lattice
vectors $\{R\}$. In the Appendix, we show that matrices of
the form given by eq.~(\ref{matrix1d:eq}) above have a highly
degenerate eigenvalue spectrum and, in particular, one
eigenvalue $\lambda_1 = A(k)+(n_c-1)B(k)$ and $n_c-1$ degenerate
eigenvalues all equal to $\lambda_2 = A(k)-B(k)$. Introducing
the specific expressions for $A(k)$ and $B(k)$ from 
eqs.~(\ref{a1d:eq}) and (\ref{b1d:eq}), we obtain the
phonon frequencies of the one-dimensional cluster crystal as:
\begin{eqnarray}
\omega_1^2(k) & = & \frac{n_c}{m}\sum_{R}v''(R)[1-\cos (kR)],
\label{ac1d:eq}
\\
\omega_i^2(k) & = & \frac{n_c}{m}\sum_{R}v''(R),\qquad\qquad i=2,3,\ldots,n_c.
\label{opt1d:eq}
\end{eqnarray}

The first branch, eq.~(\ref{ac1d:eq}), is the acoustical one, whereas
the remaining, $n_c-1$ branches are not only all degenerate but they
are also $k$-independent as well. This is a remarkable result that shows
that cluster crystals are physical realizations of the Einstein phonon
model. In addition, we observe that the squared frequencies scale linearly 
with the occupancy, $n_c$. The dependence on the density $\rho$
of the crystal comes through the fact that the lattice vectors $\{R\}$,
measured in units of the interaction range $\sigma$, 
change with concentration. We can thus write down the scaling
relations for the acoustical
and optical phonon frequencies, $\omega_{\rm ac}(k;n_c,\rho)$ and
$\omega_{\rm op}(k;n_c,\rho)$ respectively,
their dependence on the occupation number $n_c$ and the density
$\rho$ of the crystal:
\begin{eqnarray}
\omega_{\rm ac}(k;n_c,n_c\rho) & = & \sqrt{n_c}\,\omega_{\rm ac}(k;n_c=1,\rho),\qquad\qquad\,
(n_c \geq 1)
\label{scaleac:eq}
\\
\omega_{\rm op}(k;n_c,n_c\rho/2) & = & \sqrt{\frac{n_c}{2}}\,\omega_{\rm ac}(k;n_c=2,\rho),
\qquad\qquad
(n_c \geq 2)
\label{scaleop:eq}
\end{eqnarray}

As shown in sec.~\ref{zerot:sec}, the dependence of the lattice constant
on density 
is rather weak and becomes increasingly suppressed
as the density grows, as long as we stay within the limits of 
thermodynamic stability of the cluster crystals.
Concomitantly, to an approximation that becomes
more and more accurate as the density grows, we can state that having
calculated the phonon spectra for some $n_c \gg 1$ is sufficient to accurately
predict the same for all higher values of $n_c$.
Essentially, a single phonon spectrum for some lattice constant 
in the middle of the stability domain of the highly occupied crystal
is, to zeroth-order approximation, the same for all densities in that
regime and in going from the $n_c$- to the $n_c+1$-occupied crystal,
one can simply take the frequencies and multiply them by the factor
$\sqrt{(n_c+1)/n_c}$ to obtain the new ones. 

All these results carry over to arbitrary dimensions. The dynamical matrix
of eq.~(\ref{matrix1d:eq}) takes in $d$-dimensions a similar form, 
however the entries $A(k)$ and $B(k)$ become themselves $d\times d$
matrices ${\bf A}({\bf k})$ and ${\bf B}({\bf k})$ with entries:
\begin{eqnarray}
A_{\alpha\beta}({\bf k}) & = & \sum_{\bf R}
\frac{\partial^2 v({\bf R})}{\partial R_{\alpha}\partial R_{\beta}}
[n_c - \cos ({\bf k}\cdot{\bf R})],
\label{aalphabeta:eq}
\\
B_{\alpha\beta}({\bf k}) & = & -\sum_{\bf R}
\frac{\partial^2 v({\bf R})}{\partial R_{\alpha}\partial R_{\beta}}
\cos ({\bf k}\cdot{\bf R}),
\label{balphabeta:eq}
\end{eqnarray}
where $\alpha,\beta$ are Cartesian coordinates and the summations run
over all Bravais lattice vectors $\{{\bf R}\}$. The eigenvalue spectrum
of this matrix is, as in the one-dimensional case, highly degenerate:
a number $d(n_c - 1)$ of them, corresponding to the optical branches,
are given by the difference of the diagonal elements of the block
submatrices ${\bf A}({\bf k})$ and ${\bf B}({\bf k})$.  
For crystals of cubic symmetry, as is our case, we make use 
of the identity  
$\partial^2/\partial R_{\alpha}^2 = (1/d)\nabla^2$ for each Cartesian
coordinate $\alpha=1,\ldots,d$, 
to find that
the frequencies of the optical branches, $\omega_{\rm op}$,
are given by the simple expression:
\begin{equation}
\omega^2_{\rm op} = \frac{n_c}{md}\sum_{\bf R} \nabla^2 \phi({\bf R}),
\label{einstein:eq}
\end{equation}
where, as in the one-dimensional case, the frequencies are ${\bf k}$-independent
and we obtain, once more, Einstein crystals. In fact, all
optical modes are internal, ``breathing modes'' of the individual
clusters on the lattice sites. In the acoustical oscillation modes,
on the other side, all
the particles within a given site move together as a composite
object, i.e., the instantaneous displacements of all particles
in a cluster are identical. 

\begin{figure}
\begin{center}
\includegraphics[width=12cm,clip=true]{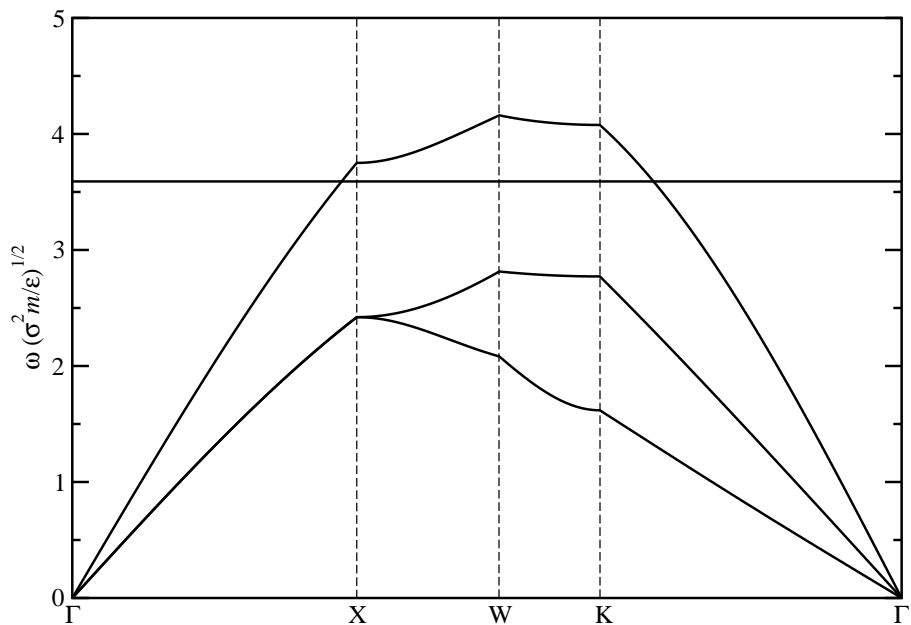}
\end{center}
\caption{The phonon spectrum of the double occupied fcc GEM4-crystal
at density $\rho^* = 1.0$. The thick vertical line is the triply degenerate
optical branch.}
\label{phonons_2d_double_stable:fig}
\end{figure}

The result of 
eq.~(\ref{einstein:eq}) above, has been anticipated in ref.~\cite{likos:jcp:07}
at the level of 
an approximation. There, the limit $n_c \gg 1$ has been considered,
and the lattice oscillations have been modeled by deviations of a 
single particle from each of the lattice sites, while the remaining $n_c - 1$ ones
remain approximately immobile, due to the large mass discrepancy between
the two entities. That simplified
approach results into an expression almost identical
to eq.~(\ref{einstein:eq}), with $n_c$ being replaced by $n_c - 1$,
the two quantities becoming arbitrarily similar in the
region $n_c \gg 1$.
In the same limit, it has been found that density functional theory
predicts precisely the same dependence of the localization parameter
of the one-particle density around the lattice sites. The phonon calculation
performed here shows that all these features stem from the peculiar,
Einstein-solid phonon spectrum of cluster crystals. 
In Fig.~\ref{phonons_2d_double_stable:fig},
we show as a concrete example the phonon spectrum of a doubly occupied
GEM4 fcc crystal, featuring the acoustical and the triply degenerate 
optical branches.

\subsection{Mechanical stability}
\label{stability:sec}

We return now to the acoustical modes, to establish the existence
of a cascade of mechanical instabilities in the system. Due to the
inversion symmetry of the Bravais lattice, the acoustical modes
of the crystal have low-$k$ expansion that involves only even powers
of the wavenumber:
\begin{equation}
\omega^2_{\rm ac}({\bf k}) = 
c_2({\bf \hat k};\rho,s)k^2 + c_4({\bf \hat k};\rho,s)k^4 + O(k^6),
\label{expand:eq}
\end{equation}
with coefficients $c_i({\bf \hat k};\rho,s)$ 
that depend on the density $\rho$ as well as
the propagation direction ${\bf \hat k}$ and the branch 
index $s$.\footnote{Note that the quantity $c_2({\bf \hat k};\rho,s)$
is the squared speed of sound of the $s$-branch for propagation along the 
${\bf \hat k}$ direction. For completeness,
we also note that an expansion of the 
form of eq.~(\ref{expand:eq})
holds true for the optical modes
as well, with the addition of a term $c_0({\bf \hat k};\rho,s)$.} 
Stability against long-wavelength modulations means that the
coefficient of the quadratic term, $c_2({\bf \hat k};\rho,s)$,
must be positive for all directions and
branches at a given density. Accordingly,
a mechanical instability occurs at the point in which $c_2$ changes sign,
the direction ${\bf \hat k}$ and the branch $s$ for which this
first occurs, as density grows, being the most unstable ones.

\begin{figure}
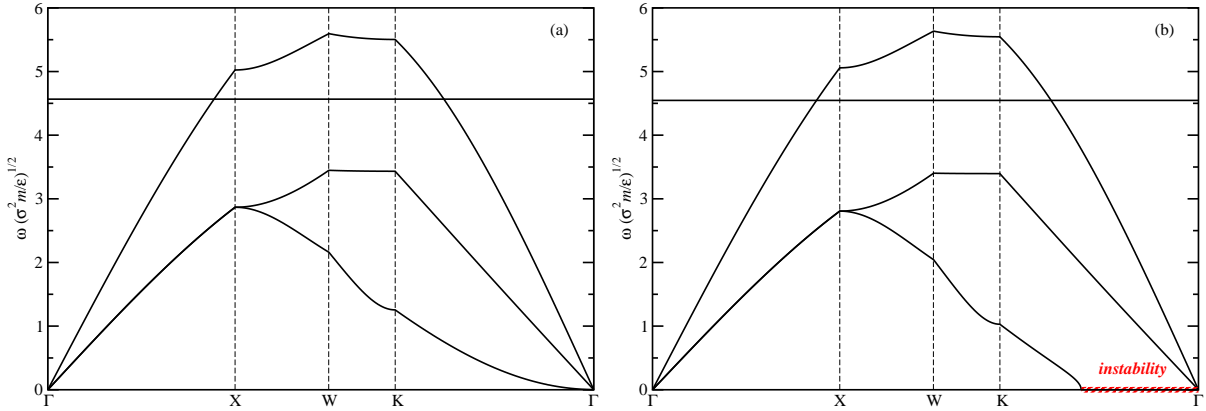

   \begin{center}
   \begin{minipage}[b]{16.0cm}
         \includegraphics[width=7.9cm,clip]
         {./fig8a.eps}
         \includegraphics[width=7.9cm,clip]
         {./fig8b.eps}
   \end{minipage}
   \end{center}
\caption{(a) The phonon spectrum of the fcc GEM4 double occupied
crystal at the threshold density $\rho_{\times,2}^* = 1.441$. Note
the dependence $\omega(k) \propto k^2$ of the lowest-lying acoustical branch
along the $\Gamma{\rm K}$ direction.
(b) The same at a density beyond the threshold value,
$\rho^* = 1.5$, where a whole region, marked by the cross-hatched
symbols, has become unstable.}
\label{doubleinstability:fig}
\end{figure}

In sec.~\ref{acoustical:sec} we established the existence of a threshold
density $\rho_{\times,1}^*$ for the single-occupied crystal, at which
a mechanical instability occurs along the $\Gamma {\rm K}$ direction
of the fcc GEM4-crystal; evidently, 
it holds $c_2({\bf \hat k};\rho_{\times,1}^*,s) = 0$
for the lowest branch in the $\Gamma {\rm K}$-direction ${\bf \hat k}$ there.
This fact, combined with the scaling relation,
eq.~(\ref{scaleac:eq}), implies that there exist infinitely many mechanical
instabilities, one for each occupancy $n_c$, and occurring at the
corresponding threshold densities $\rho_{\times,n_c}^*$ given by:
\begin{equation}
\rho_{\times,n_c}^* = n_c\rho_{\times,1}^*.
\label{rhoxnc:eq}
\end{equation} 
We have numerically confirmed this prediction for GEM4-crystals of a few 
low occupancies and we show, as a representative result, in 
Fig.~\ref{doubleinstability:fig} the phonon spectra of double occupied GEM4-crystals exactly
at the threshold density $\rho_{\times,2}^*$, and at a slightly higher one
where a number of modes feature imaginary frequencies.

\begin{figure}
\begin{center}
\includegraphics[width=12cm,clip=true]{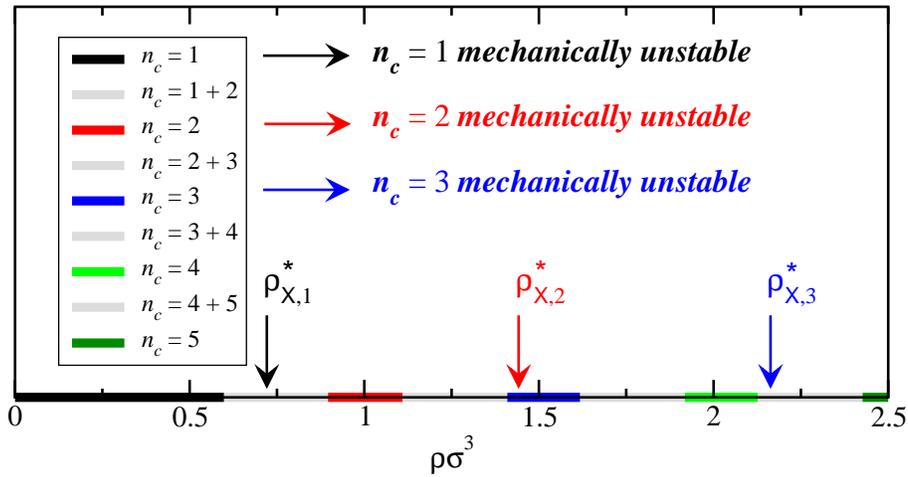}
\end{center}
\caption{The regions of stability of the pure, $n_c$-occupied
GEM4-crystals at $T=0$, as indicated in the legend, are shown 
together with the coexistence regions (marked gray) and the 
threshold densities $\rho_{\times, n_c}^*$ of mechanical stability
marked by the arrows. The latter are color-coded in the same way
as the thermodynamic stability regions of the $n_c$-occupied
crystals.}
\label{stability_diagram:fig}
\end{figure}

In Fig.~\ref{stability_diagram:fig} we graphically summarize the 
zero-temperature thermodynamic and stability phase diagrams for the
GEM4-model, showing the sequence of the first few occupancies for
the fcc crystals. It can be seen that the instability densities
$\rho_{\times,n_c}^*$ of the $n_c$ occupied crystals occur well
beyond the limits of their thermodynamic stability. The existence
of the former can be understood in terms of the particular form
of the interaction potential. Suppose we restrict ourselves to
single occupied crystals and we increase the density. This compresses
the system, causing the lattice constant to shrink. Since the
interaction potential contains a flat part at small separations,
and in view of the fact that the restoring forces in the crystal come
mainly from the nearest neighbors, at some threshold value of the
shrinking lattice constant the restoring force will vanish, at least
along one crystallographic direction (in our case, along the
nearest-neighbor vector). This signals the mechanical
instability. Now, {\it the same} argument carries over to a 
crystal of occupancy $n_c$ at $n_c$ times the threshold density
of the single occupied crystal, since the two are copies of each
other as far as the lattice constant is concerned and only the
forces in the latter are $n_c$ times bigger than the ones in the
former. If the restoring force in some direction of the single
occupied crystal vanishes, it will also do so at the same direction
for the $n_c$ occupied crystal. We see, therefore, that the whole
cascade of isostructural transitions is a straightforward consequence
of the transition from the single to the double occupancy, which can
be seen as the strategy of the system to avoid the mechanical collapse.
Indeed, multiple occupancy allows the system to readjust its lattice
constant and to arrange the particles in such a way that all
restoring forces are nonvanishing, offering the solid stability
against thermal fluctuations. The fact that the thermodynamic phase
transitions preempt the mechanical ones is reassuring and in line
with a number of other examples, in which a system undergoes a first-order
transition that has an instability line deep in the region of the
two-phase coexistence (or even beyond it, as in our case). The existence
of the instability, however, has implications on the nucleation rates
of the multiply occupied crystals, since, as one approaches it, it becomes
exceedingly difficult to maintain thermodynamically metastable crystals
of the wrong occupancy, a feature that should drastically suppress the
corresponding nucleation barriers.

\section{Phonon thermodynamics and isostructural critical points}
\label{phdg:sec}

Up to now our considerations were limited to $T=0$ and to mechanical
equations of motion of the particles around their equilibrium positions.
Here we will reintroduce finite temperatures and calculate free energies
and the ensuing phase diagrams that follow on the basis of phonon
modes {\it alone}. 

The normal modes render the harmonic part of the
system's Hamiltonian diagonal, i.e., a decoupling among the modes
results. In a crystal with $N$ particles, occupancy per site $n_c$ and
$N_{\ell}$ lattice sites, we have $N = n_c N_{\ell}$. The original
coordinates are the $3N$ Cartesian displacements of the particles
from their equilibrium sites. In the normal mode description, these
are replaced by $3n_c$ branches, each one containing $N_{\ell}$
values of the wavevector ${\bf k}$ compatible with the Born-von Karman
boundary conditions, giving rise, consistently, to $3n_c N_{\ell} = 3N$
coordinates, as in the original description. The Helmholtz free energy 
$F$ of the
harmonic solid takes, after the introduction of the normal modes, 
and on the basis of the decoupling among them, the simple form:
\begin{equation}
\frac{F}{V} = \frac{U}{V} + \frac{k_{\rm B}T}{(2\pi)^3}\sum_{s=1}^{3n_c}
\int^{'}{\rm d}^3k\ln\left[\frac{\hbar\omega_s({\bf k})}{k_{\rm B}T}\right],
\label{fren:eq}
\end{equation}
which includes the contributions from the momentum degrees of freedom
as well as the zero-point energy $U$ (the lattice sum) of the crystal.
In eq.~(\ref{fren:eq}) above, the integral over $k$ extends within the
first Brillouin zone of the crystal, a fact indicated by the prime,
whereas the sum runs over the $3n_c$ branches of the phonon spectrum,
characterized by the dispersion curves $\omega_s({\bf k})$.

Let us now focus on the limit $n_c \gg 1$. Here, the $3(n_c -1)$ optical
branches dominate over the three acoustical ones, so we ignore the 
latter
in what 
follows,\footnote{The acoustical branches can also be taken into
account analytically, within the Debye approximation.}
and we further make the approximation $n_c - 1 \cong n_c$. Since
the optical frequencies $\omega_{\rm op}$ are 
${\bf k}$-independent, we can pull the whole integrand out of
the integral in eq.~(\ref{fren:eq}) above. The remaining integral
$(2\pi)^{-3}\int^{'}{\rm d}^3k$ yields the density of {\it lattice
sites}, $N_{\ell}/V$, whilst the summation over $s$ gives the
number $3n_c$ of phonon branches; the two combine into $3n_cN_{\ell}/V
= 3N/V = 3\rho$, to give, within the stated
approximations:
\begin{equation}
\frac{F}{V} = \frac{U}{V} + 3{k_{\rm B}T}\rho
\ln\left[\frac{\hbar\omega_{\rm op}}{k_{\rm B}T}\right].
\label{fren1:eq}
\end{equation}

Apart from the manifest $\rho$-dependence of the second term on
the right-hand-side of eq.~(\ref{fren1:eq}), there is an implicit
one through the optical frequency $\omega_{\rm op}$; an explicit
expression for the latter is given in eq.~(\ref{einstein:eq}).
The density enters there through the dependence of $n_c$ and the
lattice vectors $\{{\bf R}\}$ on $\rho$. As discussed in length
in sec.~\ref{zerot:sec}, at the limit $n_c \gg 1$ considered here,
the lattice constant can be taken as density independent, due to
the approximate property $n_c \propto \rho$. Accordingly, we 
define the $\rho$-independent frequency $\omega_0^2 = \sum_{\bf R}\nabla^2\phi({\bf R})/(3m)$,
for which the set of lattice vectors $\{{\bf R}\}$ is determined by the
interparticle interaction {\it alone} and, in particular, by the
property that the length of the shortest reciprocal lattice vector of the crystal
coincides with $k_*$, the wavenumber at which $\tilde v(k)$ has its most
negative amplitude \cite{likos:jcp:07}. Then, we have 
$\omega_{\rm op} \cong \sqrt{n_c}\omega_0$ and its $\rho$-dependence
comes exclusively through $n_c$, for which we know that $n_c \propto \rho$.
Gathering all the results together, introducing them into eq.~(\ref{fren1:eq}),
and absorbing all terms linear in $\rho$ into a separate term with
some proportionality constant $C$, we arrive at the result:
\begin{equation}
\frac{F}{V} \cong \frac{U}{V} + \frac{3k_{\rm B}T}{2}\rho\ln(\rho\sigma^3) + C\rho,
\label{fren2:eq}
\end{equation}   
where we have introduced an arbitrary length scale $\sigma$ in the logarithm to
render its argument dimensionless. The last term, $C\rho$, has no
effect on the determination of phase boundaries; it acts simply
as a constant shift to the chemical potential, leaving the
pressure invariant, thus it will be dropped in what follows.
We arrive, thus, at a remarkably simple
(albeit approximate) result, eq.~(\ref{fren2:eq}), which states that the 
phonon free energy
of the cluster crystals can be expressed as the sum of the ground state
energy $U$ plus an ideal gas contribution, the latter corresponding to
a `fictitious temperature' equal to three-halves the real one.\footnote{In
$d$ spatial dimensions this would be $d$-halves the real temperature.}

\begin{figure}
\begin{center}
\includegraphics[width=12cm,clip=true]{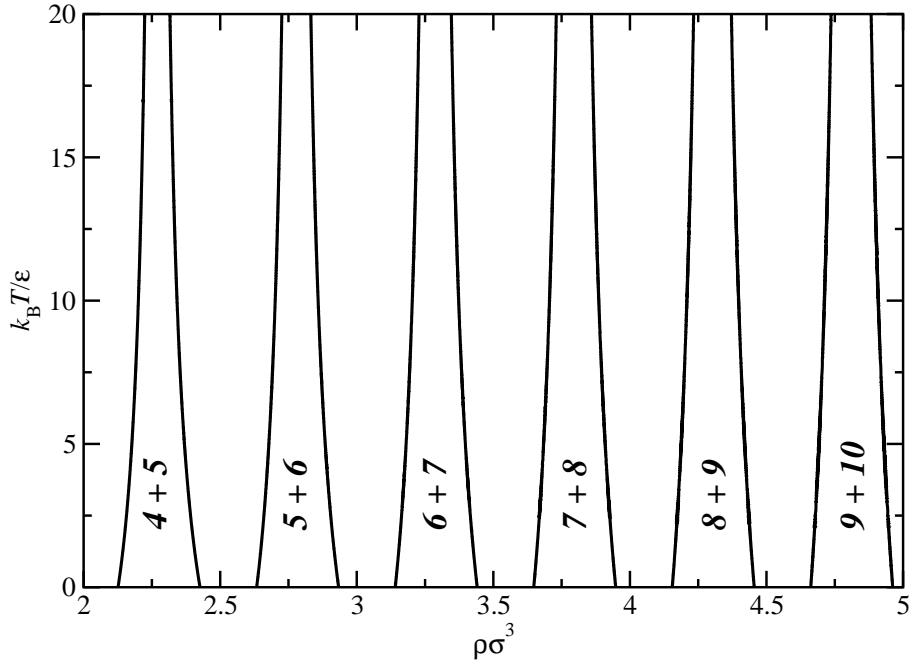}
\end{center}
\caption{The phonon-based phase diagram of the GEM4-model,
shown here, in part, for the region of moderate occupancies,
for which the theoretical approximations discussed in the
text are becoming increasingly valid. Two-phase regions of
coexistence between fcc crystals of occupancy $n_c$ and
$n_c + 1$ are indicated by the inscriptions, whereas unlabeled
domains correspond to single-phase regions.
The coexistence density
gaps found at $T=0$ are being narrowed as $T$ grows, due to
phonon contributions, but they never close. In the true phase diagram,
hopping processes eliminate regions of stability of crystals with
different site occupancies
at sufficiently
high temperatures.}
\label{isostructural:fig}
\end{figure}

We can now, on the basis of the free energy of eq.~(\ref{fren2:eq}) above
draw a {\it phononic} phase diagram of the crystal states of the system.
The qualification pertains to the fact that in writing down
eq.~(\ref{fren2:eq}), the contributions from the phonons {\it only}
have been taken into account. Both the disturbances of the phonon
spectra arising from imperfect crystals with nonuniform site occupancies
and the entropy associated with hopping are {\it not} included in
the free energy of eq.~(\ref{fren2:eq}). The first therm there simply 
gives the $T=0$ phase diagram with the phase coexistence
regions discussed in sec.~\ref{zerot:sec}. The second term, seen as 
a function of $\rho$, has curvature $3k_{\rm B}T/(2\rho) > 0$, thus
it will act to narrow the width of the density gaps as temperature
grows. 

The phase diagram can be drawn by performing 
the double tangent construction on the $F/V$ vs.~$\rho$-curves
and it is shown in Fig.~\ref{isostructural:fig}
for a region of moderate values of the density and thus site occupation
numbers. As anticipated above, the phonon contributions lead to a 
narrowing of the density gaps as temperature increases. Nevertheless,
the crystals of different occupancy remain distinct at all temperatures,
as is natural for an approach that does not allow hopping and smooth
adjustments of the lattice constant. The phonon-based phase diagram
of Fig.~\ref{isostructural:fig} must now be juxtaposed to the
`exact' one, arising from 
density functional theory \cite{likos:jcp:07,mladek:prl:06,
mladek:jpc:07} and computer simulations \cite{daan:prl:07}. The full
phase diagram is free of the coexistence regions between crystals
of different occupancy, for temperatures at least as high as 
$T^* = 0.1$. Evidently, all the contributions from hopping,
inhomogeneous lattice site occupancy and anharmonicities, bring
about the effect of `washing out' the step-like behavior of $n_c$
and thus eliminating the first-order phase transitions between
crystals of different site occupancies at sufficiently high
temperatures. On the other hand, at $T=0$ the density gaps and
the transitions between such crystals are definitely present. 
Whichever contribution hopping and anharmonicities have on the
free energy, it cannot simply eliminate the gaps present at $T=0$
for arbitrarily small temperatures. It follows that the only
possibility to reconcile the low- and high-temperature properties
of the model is to put forward the conjecture that the coexistence
gaps between crystals of occupancy $n_c$ and crystals of 
occupancy $n_c+1$ must narrow down as $T$ grows and terminate
at critical points. As there are infinitely many gaps at $T=0$,
it follows that the system must feature infinitely many, low-$T$
critical points at which the isostructural fcc-fcc transitions
end. Preliminary simulations seem to confirm 
this hypothesis \cite{bianca:prl:11}, 
whereby the critical temperature $T_c$ has been estimated in the 
order of magnitude
$k_{\rm B}T_c/\varepsilon \sim 10^{-2}$.

The isostructural fcc-to-fcc transitions and their termination 
at a critical point bear some similarities to the ones discovered,
and extensively investigated, in the 1990s for hard-sphere systems with
short-range attractive or repulsive 
interactions \cite{young:jcp:80,daan:prl:94,daan:pre:94,likos:jpcm:94,
rascon:prb:95,rascon:prl:95,likos:senatore:95,
likos:nemeth:95,rascon:jcp:95,lang:jpcm:99}. 
Also in that
case, two fcc-lattices with different values of the lattice constant
were found to coexist at sufficiently low temperatures, the 
difference between the two gradually disappearing as the temperature
grows. There are, however, a number of important differences. First,
in the previous cases, the coexistence was not a ground-state
feature, since it disappeared below the triple, fcc-fcc-gas temperature,
to give its place to a usual, fcc-gas coexistence. Second, the
two crystals had the same occupancy there, unity, whereas here
they are characterized by different values of $n_c$. Finally,
the systems at hand present {\it infinitely many} such regions
of coexistence and corresponding critical temperatures, as opposed
to a single one in previous models.
 
\section{Summary and conclusions}
\label{concl:sec}

We have investigated the ground states and the phonon spectra
of cluster crystals formed by $Q^{\pm}$-particles, finding 
a cascade of isostructural transitions and the emergence of these
systems as realizations of Einstein crystals. We have put
forward the hypothesis that the systems display infinitely many,
low-temperature critical points in which the fcc-fcc transitions
terminate.
The predictions made here
hold quite generally for all $Q^{\pm}$ potentials, {\it provided}
there exist mutual, nonvanishing forces between the particles,
so that the arguments presented in sec.~\ref{zerot:sec} hold.
An important exception is the so-called 
penetrable sphere model (PSM) \cite{likos:psm:98},
which is 
formally a GEM with infinite exponent, and for which no forces act
between the particles; consequently, neither the ground state
requires a vanishing force on each particle due to its neighbors nor
is a harmonic expansion of the potential possible. For the PSM,
the low-temperature phase diagram features {\it gradual} crossovers
from low- to high-occupancy fcc-crystals instead \cite{likos:psm:98}.

Finally, we comment on the possible limitations of our simplifying
assumption that in any given solid at $T=0$ all sites have the
same, integer occupancy $n_c$. As discussed in sec.~\ref{zerot:sec},
the occupancy can be at most a rational number. Therefore, even in
full generality, isostructural transitions between rationally
occupied crystals would appear -- the rational numbers forming
a discrete set, as opposed to the general, real occupancy found
by density functional theory and simulation at moderate and high
temperatures. It follows that the conclusions on the necessity
to terminate the isostructural transitions at critical points remain
unaffected.
 
\ack The authors acknowledge useful discussions with Ronald Blaak,
Bianca Mladek, Daniele Coslovich, Lukas Strauss, Gerhard Kahl, and Daan Frenkel. 

\section*{Appendix}

Consider a $\nu \times \nu$ Toeplitz matrix ${\mathcal{M}}_{\nu}$ with the
property that all its diagonal elements are equal to $A$ and the
nondiagonal ones equal to $B$, i.e.:
\begin{equation}
{\mathcal{M}}_{\nu}  = \left({\begin{array}{ccccccc}
A & B & B & . & . & . & B\\
B & A & B & . & . & . & B\\
B & B & A & . & . & . & B\\
B &   &   & . &   &   & B\\
B &   &   &   & . &   & B\\
B &   &   &   &   & . & B\\
B & B & B & . & . & . & A\end{array} } \right), 
\end{equation}
Replacing the
first diagonal element of ${\mathcal{M}}_{\nu}$ with $B$ we obtain the
matrix ${\bar \mathcal{M}}_{\nu}$ with the form:
\begin{equation}
{\bar \mathcal{M}}_{\nu}  = \left({\begin{array}{ccccccc}
B & B & B & . & . & . & B\\
B & A & B & . & . & . & B\\
B & B & A & . & . & . & B\\
B &   &   & . &   &   & B\\
B &   &   &   & . &   & B\\
B &   &   &   &   & . & B\\
B & B & B & . & . & . & A\end{array} } \right),
\end{equation}
The following statements hold true for the determinants 
$|{\mathcal{M}}_{\nu}|$ and $|{\bar \mathcal{M}}_{\nu}|$ of the two matrices:
\begin{eqnarray}
\label{mnu:eq}
|{\mathcal{M}}_{\nu}| & = & (A-B)^{\nu - 1}[A + (\nu - 1)B];\\
\label{mbnu:eq}
|{\bar \mathcal{M}}_{\nu}| & = & B(A-B)^{\nu - 1}.
\end{eqnarray}
The proof follows by induction. Evidently, the relations hold for $\nu = 2$.
Now assume that they hold for $\nu = n$. Developing the determinant of the matrices
${\mathcal M}_{n + 1}$ and ${\bar \mathcal M}_{n + 1}$ around the entries
of their first rows, we obtain:
\begin{eqnarray}
\label{mnupone:eq}
|{\mathcal{M}}_{n+1}| & = & A|{\mathcal{M}}_{n}| 
                                - n B |{\bar \mathcal{M}}_{n}|;\\
\label{mbnupone:eq}
|{\bar \mathcal{M}}_{n+1}| & = & B |{\mathcal{M}}_{n}|
                                - n B |{\bar \mathcal{M}}_{n}|.
\end{eqnarray}
Introducing into eqs.~(\ref{mnupone:eq}) and (\ref{mbnupone:eq}) 
the expressions of eqs.~(\ref{mnu:eq}) and (\ref{mbnu:eq}),
assumed valid for $\nu = n$, results into the relations, now proven
valid for $\nu = n+1$: 
\begin{eqnarray}
|{\mathcal{M}}_{n+1}| & = & (A-B)^{n}(A + n B);\\
|{\bar \mathcal{M}}_{n+1}| & = & B(A-B)^{n},
\end{eqnarray}
which completes the proof.

The characteristic polynomial $p_{\nu}(\lambda)$ 
of the matrix ${\mathcal{M}}_{\nu}$ is, due to the special form of the
latter, obtained through the formal substitution 
$A \to A - \lambda$ in 
the expression for $|{\mathcal{M}}_{\nu}|$. Accordingly, the
eigenvalue equation $p_{\nu}(\lambda) = 0$ reads, for this matrix, as:
\begin{equation}
(A-B-\lambda)^{\nu - 1}[A + (\nu - 1)B-\lambda] = 0,
\end{equation}
from which the eigenvalue spectrum 
of the matrix ${\mathcal{M}}_{\nu}$ follows as:
\begin{eqnarray}
\lambda_1 & = & A + (\nu - 1)B; \\
\lambda_i & = & A - B.\qquad\qquad\qquad (i = 2,3,\ldots,\nu)
\end{eqnarray}
The second equation manifests
the degeneracy of $\nu - 1$ eigenvalues of the matrix.

\end{document}